\documentclass[pre,reprint,twocolumn,amsmath,10pt,amssymb,aps,superscriptaddress,showkeys,showpacs]{revtex4-1}\input epsf
\usepackage{bm}
\usepackage{soul}
\usepackage{graphicx}
\usepackage{makeidx}
\usepackage{chemarr}
\usepackage{multirow}
 \usepackage{color}
 \usepackage[normalem]{ulem}
 \usepackage{appendix}
\definecolor{darkgreen}{rgb}{0,0.6,0}
 \definecolor{orange}{rgb}{0.99,0.257,0}

\def\epsilon{\varepsilon}

\def\beqr{\begin{eqnarray}}
\def\eqnr{\end{eqnarray}}
\def\beq{\begin{equation}}
\def\bc{\begin{center}}
\def\ec{\end{center}}
\def\eqn{\end{equation}}

\usepackage{amsmath}

\usepackage{graphicx}
\usepackage{dcolumn}
\usepackage{bm}

\begin{document}


\title{Percolation transition in phase separating active fluid}
\author{Monika Sanoria}
\author{Raghunath Chelakkot}
\email{raghu@phy.iitb.ac.in}
\author{Amitabha Nandi}
\email{amitabha@phy.iitb.ac.in}
\affiliation{Department of Physics, Indian Institute of Technology Bombay, Powai, Mumbai, 400076, India.}
\begin{abstract}
The motility-induced phase separation  exhibited by active particles with repulsive interactions is well known. We show that the interaction softness of active particles destabilizes the highly ordered dense phase, leading to the formation of a porous cluster which spans the system. This \emph{soft limit} can also be achieved if the particle motility is increased beyond a critical value, at which the system clearly exhibits all the characteristics of a standard percolation transition. We also show that in the athermal limit, active particles exhibit similar transitions even at low motility. With these additional new phases, the phase diagram of repulsive active particles is revealed to be richer than what was previously conceived. 
\end{abstract}

\maketitle
\section{Introduction}
 Active matter systems consist of microscopic entities that consume chemical energy and convert it to mechanical motion \cite{Ramaswamy2010,vicsek2012collective,Marchetti2013,cavagna2010scale,BECCO2008scale,zhang2010collective,beer2020phase,tan2020topological,JULICHER2007,schaller2010polar,sumino2012large,rubenstein2014programmable, KokotPNAS2017, cohenPRL2014,gompper2020, bricard2013emergence, thutupalli2011swarming,Buttinoni2013}. Such systems display a variety of collective properties as a result of the interplay between the activity and the nature of interactions between the constituent entities~\cite{Buttinoni2013,vanderLinden2019,wysocki2014cooperative,Cates2015,digregorio2018full,caporusso2020motility,stenhammar2013continuum,levis2017active,klamser2018thermodynamic,mandal2019motility,caprini2020spontaneous,Das2020,Lee_2013,Fily2014,Elgeti_2013,shi2020self,Su_2021,cates,stenhammar2014phase}. Minimal theoretical models used to study the collective behaviour of active matter belongs to two major categories. The first one involves purely alignment interaction between the active polar entities and leads to an emergence of global polar order ~\cite{shaebani2020computational,Viscek1995,vicsek2012collective,peruani2011traffic,ginelli2010relevance,chate2008modeling,chate2006simple}. The second class of models study the effect purely excluded volume interactions between the entities leading to a phase separated state \cite{Fily2012,Redner2013,catesReview2015,stenhammar2014phase,digregorio2018full}, known as motility-induced phase separation (MIPS). Unlike the systems with alignment interaction, the structural properties of the phase separating systems have been so far known to be relatively simpler.  
 
 
 
 MIPS has received much attention during the last decade and have been studied extensively using particle-based simulations~\cite{Redner2013,Fily2012,digregorio2018full} and continuum descriptions~\cite{Cates2015,stenhammar2014phase,speck2015dynamical}. The generic phase behavior for nearly hard particles with short-range repulsive interactions has been explored in some detail~\cite{Redner2013,Fily2012,digregorio2018full}. However, many of the dynamical properties, especially in the dense phase, are still being revealed~\cite{caprini2020spontaneous, caporusso2020motility, shi2020self}. Also, subtle modifications in the surrounding environment and the inter-particle interactions lead to a range of intriguing collective properties~\cite{redner2013reentrant,paliwal2017non,Das2020,soker2021activity,huang2020dynamical,jose2021phase,hiraiwa2020dynamic}, thus making such phase separating systems still an interesting topic of research.

 Numerical studies on MIPS have focused on systems where overlap distance between the particles are small but non-zero~\cite{redner2013reentrant,digregorio2018full}. In such systems, an increase in motility causes a more compact hexatic packing of particles within the clusters, stabilizing the dense phase. On the other hand, it has also been shown that a higher degree of particle overlap leads to a destruction of MIPS as the particles can pass through each other~\cite{fily2014freezing}. But, there lies an interesting regime where the overlap is significant while the particles still don't pass through. Such systems have been studied earlier in the equilibrium-limit and were observed that a change in particle softness affects the high density crystalline order ~\cite{agrawal1995solid,rey1998influence}. Recent experiments with 2D passive colloids have revealed that an increase in the particle softness causes complex self assembled structures \cite{menath2021defined}. A natural question thus arises, whether such emergent complex structures are also possible for systems with soft active particles. This soft limit is  particularly relevant since many active systems occurring in nature consist of deformable entities, for example, cells in tissues.
Furthermore, since the numerical studies on phase separation have used soft particle models, it would be interesting to study the high motility \emph{soft} limit where the interparticle distances are small, to explore the possibilities of new phases beyond MIPS, and to obtain a complete understanding of their phase behaviour. 

Here, we show that {for a given particle softness,} there exists a critical value of motility for soft active particles, beyond which the structurally ordered high density cluster in MIPS becomes unstable, leading to the formation of porous, connected clusters which spans the system size. Our detailed analysis reveals that the transition to this connected state {is caused by an interplay between motility and particle softness and} shows all the characteristics of a standard percolation transition.

\noindent
\section{Model}
Our numerical model consists of $N$ active Brownian particles (ABPs), at position $r_i$ with the direction of self-propulsion $\hat{\mathbf{n}}_i=(\cos{\theta_i},\sin{\theta_i})$ respectively, confined in 2D periodic boundary system of box length $L$. The dynamical equations of motion are given by the overdamped Langevin equation as \cite{Fily2012}
\beqr
\label{eq1}
{\dot{\bf r}_i} &=& \mu~\sum_j{\bf F}({\bf r}_{ij})+ v_p~\hat{\mathbf{n}}_i,\\
\dot{\theta}_i &=& {\xi}_i^R.
\label{eq2}
\eqnr
Here $v_p$ is the self-propulsion speed and $\mu$ is the mobility. The Gaussian white noise $\xi_i^R$ with mean zero and variance $2D_r$ satisfies the relation $\left<\xi^R_i(t)~\xi^R_j(t')\right> = 2~D_r \delta_{ij}~\delta(t-t')$, where $D_r$ is the rotational diffusion coefficient. The repulsive force between a pair of particles is given as ${\bf F}({\bf r}_{ij}) =  k \left( \sigma - r_{ij}\right)\hat{\mathbf{r}}_{ij} ;\hspace{0.1cm}\mbox{$r_{ij} < \sigma$}$ and zero otherwise. Here $\sigma$ is the interaction diameter of the particle, $k$ is the elastic constant that controls the stiffness of the particles, and $r_{ij}=|\mathbf{r}_{i}-\mathbf{r}_{j}|$. We use the non-dimensional number $\mbox{Pe} = \frac{v_p}{\sigma D_r}$ (P\'eclet number) as a parameter to vary particle motility. We also define $\tilde{k}=\frac{\mu k}{D_r}$ as the non-dimensional effective stiffness parameter. We keep $D_r = 0.005$, $\mu=1$, $\sigma=1$, while varying $k$ from 5 to 40. We also vary Pe from $0$ to $3000$ and $N$ from $3649$ to $233546$. The integration time-step $\Delta t=10^{-3}$.  The packing fraction $\phi = \frac{N\pi~\sigma^2}{4L^2}$ varies from 0.1 to 0.8.

\begin{figure}[h!]
\centering
\includegraphics[width=0.5\textwidth]{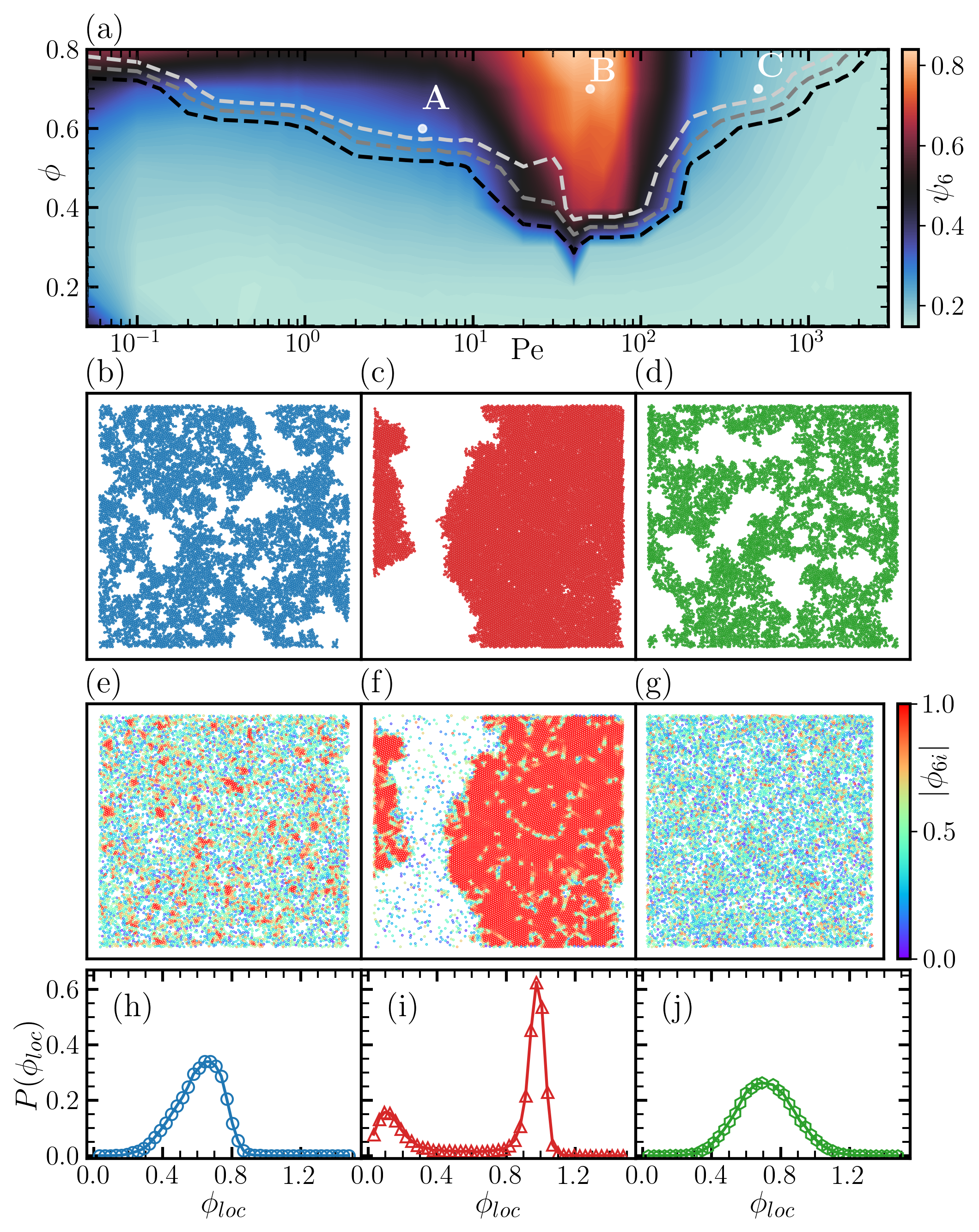}
\caption{(a) Contour map of the hexatic order parameter $\psi_6$ in the (Pe - $\phi$) plane. The dotted lines indicate the parameters for which mean largest cluster fraction $f_L$ has values 0.2 (black), 0.4 (grey), and 0.6 (white).
A$\equiv(5,0.6)$, B$\equiv(50,0.7)$, and C$\equiv(300,0.7)$ are three representative points in the (Pe - $\phi$) plane whose phase properties are further compared. (b-d) Steady-state configuration of the largest cluster at the locations A, B, and C respectively. (e-f) The corresponding steady-state snapshots for the complete system. Colour code of the particles represents the local hexatic order $\phi_{6i}$.  (h-j) Corresponding distribution of the local density $\phi_{loc}$. For our simulations we used $N=10^4$ particles. The stiffness parameter $\tilde{k}=2000$ for all the figures.
}
\label{fig1}
\end{figure}
\noindent

\section{Results}
To quantify the phase behaviour systematically, we calculate two quantities which characterize macroscopic order as a function of Pe and $\phi$ and for a fixed particle stiffness $\tilde{k}$. First we compute the global hexatic order parameter $\psi_6 = \bigg\langle \left| \frac{1}{N} \sum \limits_{i=1}^{N} \phi_{6i} \right|\bigg \rangle$. Here $\phi_{6i}=\frac{1}{N_b} \sum \limits_{j \in N_b} e^{6 i \theta_{ij}}$, quantifies the local hexatic order in the system. Here $\theta_{ij}$ is the angle between the distance vector $\mathbf{r}_{ij}$ and the reference axis, and $N_b$ is the total number of neighbors for a particle $i$ obtained by Voronoi tessellation. Next, we also calculate the largest cluster fraction $f_L = {\left<C_m \right>}/N $, where $C_m$ is the number of particles in the largest cluster. Here, a pair of particles $i$ and $j$ forms a cluster if $r_{ij}<\sigma$.
\subsection{Phase behaviour}
The results are shown in Fig.\ref{fig1} for $\tilde{k}=2000$. In the phase diagram (Fig.\ref{fig1}(a)), the color map indicate the global hexatic order ($\psi_6$). The dotted lines represent the contours for $f_L =0.2, 0.4~ \mbox{and}~ 0.6$ which demarcate the parameter range where large clusters are formed. In the case of MIPS, $\psi_6 \approx 1$ and $f_L >0.6$ as evident in Fig.\ref{fig1}(a) (for example, point B). Interestingly, there also exists a well defined region in the phase diagram with large $f_L$ and small $\psi_6$ -- for example points A and C in Fig.\ref{fig1}(a). The structural properties in these regions are quite different from MIPS as evident from Fig.\ref{fig1}(b-d), where the typical configurations of the largest cluster at the steady states are plotted. While both regions A and C show a porous structure similar to a percolated state, region B shows a single dense state as usually seen in MIPS. To further confirm the structural difference between these regions, we overlay the local hexatic order $\phi_{6i}$ on the steady state full configuration as shown in Fig.\ref{fig1}(e-g). While region B shows large $\phi_{6i}$ within the dense phase (Fig.\ref{fig1}(f)) which is a signature of MIPS, both regions A and C lack this property. However, unlike region C, region A shows small patches of large $\phi_{6i}$ values due to the formation of local clusters. {Note that the origins of these two states are different. The porous cluster seen in region A is due to an increase in the contact probability caused by low but non-zero motility and occurs with minimal particle overlap. On the otherhand, the high-motility porous clusters in region C appears after the breakdown of MIPS with enhanced particle overlap (see Fig.~5 in the Appendix-A).} This difference between A and C is further evident in Fig.\ref{fig1}(h) and (j) where we plot local density distribution $P(\phi_{loc})$. While $P(\phi_{loc})$ for both regions A and C are unimodal, the distribution in region A (see Fig.~\ref{fig1}(h)) is clearly assymmetric. This assymmetry in the distribution is a reflection of limitation of the system to form dense local regions, due to low motility. Thus, it is evident that these two porous regions are distinct at the level of local structure.  We later show that the porous structure at low Pe (Fig.\ref{fig1}(b)) is an exceptional feature and disappears in the presence of  translational thermal noise. The local density distribution for region B displays a bimodal behaviour (Fig.~\ref{fig1}(i)) as expected from a MIPS state~\cite{Redner2013}.

\begin{figure}[h!]
\centering
\includegraphics[width=0.5\textwidth]{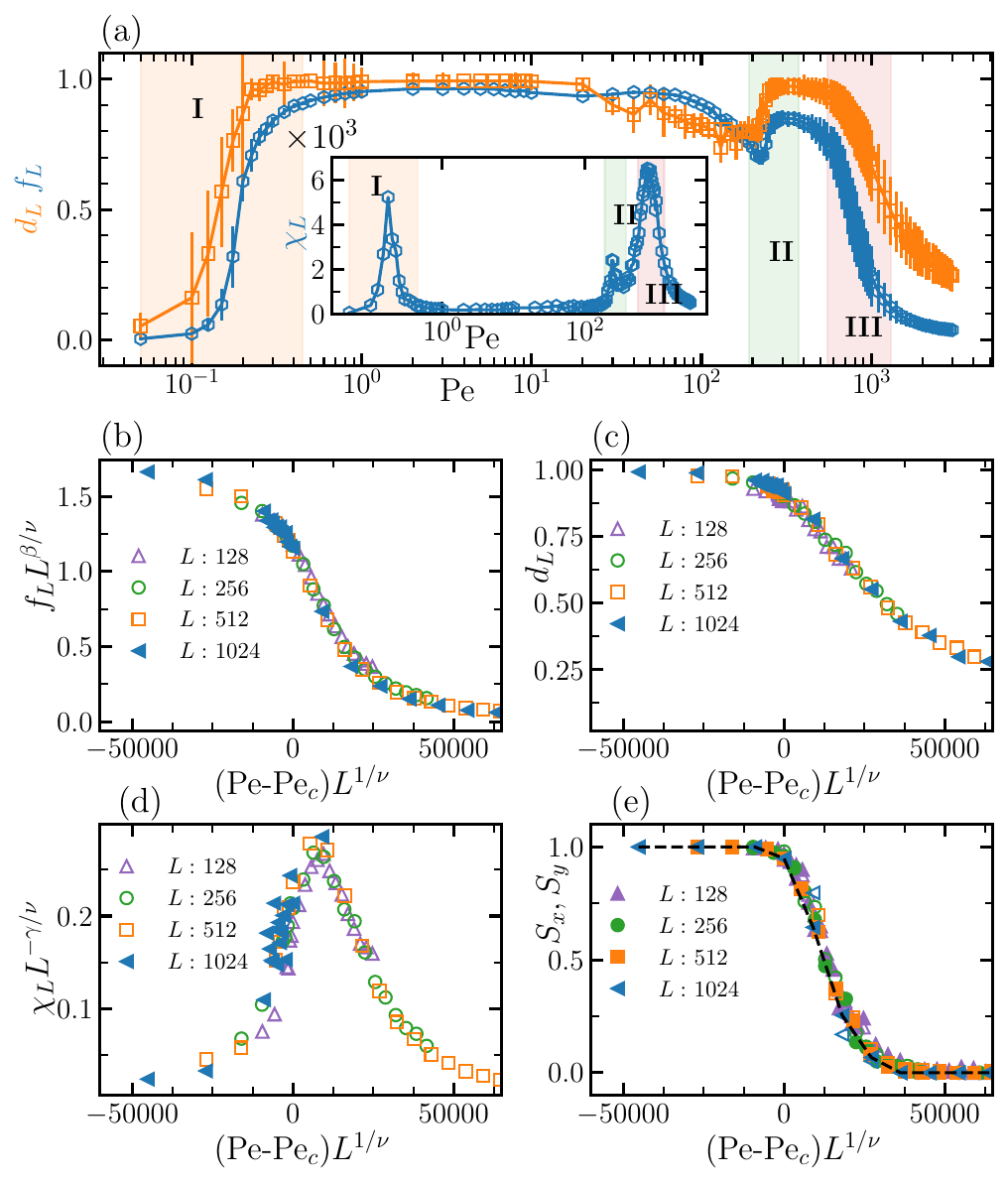}
\caption{(a) Semi-log plot of the mean largest cluster fraction $f_L$ and the average linear extension of the largest cluster $d_L$, {and susceptibility $\chi_L$ (inset)} as a function of the Pe at a constant $\phi = 0.7$. Regions I, II, and III indicate different transition regions. (b-d) showed the finite size scaling of order parameter (b)$f_L$, (c) $d_L$, (d) $\chi_L$, and (e) the spanning probabilities $S_x$ and $S_y$ with system size $L = 128, 256, 512, 1024$. The collapse occurs for exponent values $\nu \approx 1.33$, $\beta \approx 0.14$, and $\gamma \approx 2.5$. $\text{Pe}_c \approx 550$ as estimated from the fourth order Binder cumulant.}
\label{fig2}
\end{figure}
\subsection{Analysis of transitions}
It is evident from Fig.\ref{fig1} that the system goes through multiple transitions at large $\phi$ with increase in Pe. Furthermore the porous structures formed at both high and low Pe are reminiscent of a percolated state. For a closer analysis, we fix the density at $\phi = 0.7$ and study the system properties with Pe. In addition to $f_L$, we also calculate the normalized average linear extension of the largest cluster $d_L$ (see  Fig.~\ref{fig2}(a)) \cite{kyriakopoulos2019clustering}. For low Pe (region-I in Fig.\ref{fig2}(a)), both $f_L$ and $d_L$ increases from zero around Pe $\approx 0.5$, indicating the formation of a large space-filling cluster. This initial increase in both $f_L$ (and $d_L$) at small Pe is mainly due to an increase in probability of the particles to make contact with its neighbors as the particle becomes motile. However, in the presence of a translational noise, this cluster formation shifts to a higher Pe (see Fig.10 in the Appendix-C). Upon further increasing Pe, we obtain a MIPS state around Pe$\approx 20$. Note that $f_L$ does not capture this transition since its value is already large, however $d_L$ shows a decrease from its maximum value 1, since the cluster is no longer space-filling. Interestingly, at even larger Pe, (Pe $\gtrsim 100$), both $f_L$  and $d_L$ starts to decrease until they reach a minimum around Pe$\approx 200$, indicating a reduction in  cluster size. 
Both $f_L$ and $d_L$ again reaches a broad maximum around Pe $\approx 400$ beyond which it starts to decrease again. These changes in $f_L$, marked as regions II and III in Fig.\ref{fig2}(a), indicate structural transitions in the system.
To further investigate these possible transitions, we also calculate the susceptibility $\chi_L = L^2 \sigma_L$ as a function of Pe (see inset of Fig.\ref{fig2}(a)), where  $ \sigma_L = \sqrt{\left<(f_L - \left<f_L\right>^2)\right>}$ is second cumulant.  We see clear peaks in $\chi_L$ in regions-I and III, whereas in region-II the peak amplitude is relatively small. 

The peaks in $\chi_L$ is an indication of phase transitions at these regions and each of these transitions corresponds to a structural change {(see Fig.6 Appendix-B).} The first transition (Region-I) corresponds to a change from a state with $f_L \approx 0$ to another state with $f_L \approx 1$ where the system forms a porous, space-filling cluster (see Fig.~\ref{fig1}(b and e). Around Pe $\approx 20$, the system goes through a structural transition to a MIPS state (see Fig.8(a) in Appendix-B).  After the MIPS transition, there is a gradual increase in the local density within the dense cluster with Pe (see Fig.8(a) in Appendix-B), indicating an overall decrease in the cluster area. There is also a systematic increase in the number of defects and an overall decrease in the global hexatic order (see Fig.8(b-d) in Appendix-B). Around Pe $\approx 300$, the MIPS state dissolves into a porous, space-filling cluster {(Fig.8(a) in Appendix-B)}. The observed dip in $f_L$ and peak in $\chi_L$ in region-II is associated with this structural change. Further increase in Pe leads to yet another structural transition at {Pe$>1000$,} where the space-filling cluster breaks up into a large number of smaller clusters. A peak in $\chi_L$ at region-III captures this transition. To test the robustness of the transitions, we have also studied the system in the presence of translational noise of thermal origin. This is characterized an additional noise term $\xi_i(t)$ in Eq.~\ref{eq1}, such that $\langle{\xi_i(t)\xi_j(t')}\rangle=2~D \delta_{ij}~\delta(t-t')$, where $D=\frac{\sigma^2D_r}{3}$. We note that the transition captured by $f_L$ and $\chi_L$ at low Pe disappears completely, while the transition to MIPS state is now clearly identified (see Fig.10 in Appendix-C). The transitions at high Pe beyond MIPS (II and III) however remains unaffected (see Fig.10 in Appendix-C). This clearly indicates that while the transition at low Pe is also a function of temperature, the transitions seen beyond MIPS are purely dependent on the interplay of the  strength of the nonequilibrium forces and the interaction softness. 

The transition in region-III is particularly interesting as it shows the traits of a conventional percolation transition over a wide range of Pe and is robust to thermal fluctuations. To verify the existence of a percolation transition, we also calculate the spanning probability of the largest cluster, $S_x$ and $S_y$ in both $x$ and $y$ directions respectively. We further perform a finite-size scaling for $L = 128, 256, 512, \text{and}\:~1024$. For a standard percolation transition, the order parameter $f_L$ and the susceptibility $\chi_L$ should satisfy the scaling relations $f_L=L^{-\beta/\nu}f((\mbox{Pe}-\mbox{Pe}_c)L^{1/\nu})$ and $\chi_L=L^{-\gamma/\nu}g((\mbox{Pe}-\mbox{Pe}_c)L^{1/\nu})$ respectively  \cite{stauffer2018introduction}. Here Pe$_c$ is the critical parameter value at the transition point, and $\beta, \nu,$ and $\gamma$  are the universal exponents. To determine the $\mbox{Pe}_c$ we calculate the fourth order Binder cumulant {$U_L=\frac{1}{2}~(3 - \frac{\langle d_L^4 \rangle}{\langle d_L^2 \rangle^2})$} for different $L$ and find that the  crossing point is at $\mbox{Pe}_c\approx550$ {(see Fig.7 in Appendix-B)}. Using this $\mbox{Pe}_c$ we scale $f_L$, $d_L$ and $\chi_L$, and $S_{x/y}$ values and find a good collapse with $L$, for the exponents $\nu \approx 1.33$, $\beta \approx 0.14$ and $\gamma \approx 2.5$ (see Fig.\ref{fig2}b-d). These critical exponents are in good agreement with the known universal exponents for standard 2d percolation \cite{stanleyIOP1981}. {We also quantified the error in the estimation of $\mbox{Pe}_c$ value which was $\lesssim 5$\%.} Thus, the scaling of the measured quantities and the values of the critical exponents confirms that the transition in region-III is indeed a percolation transition.

\begin{figure}[h!]
\centering
\includegraphics[width=0.5\textwidth]{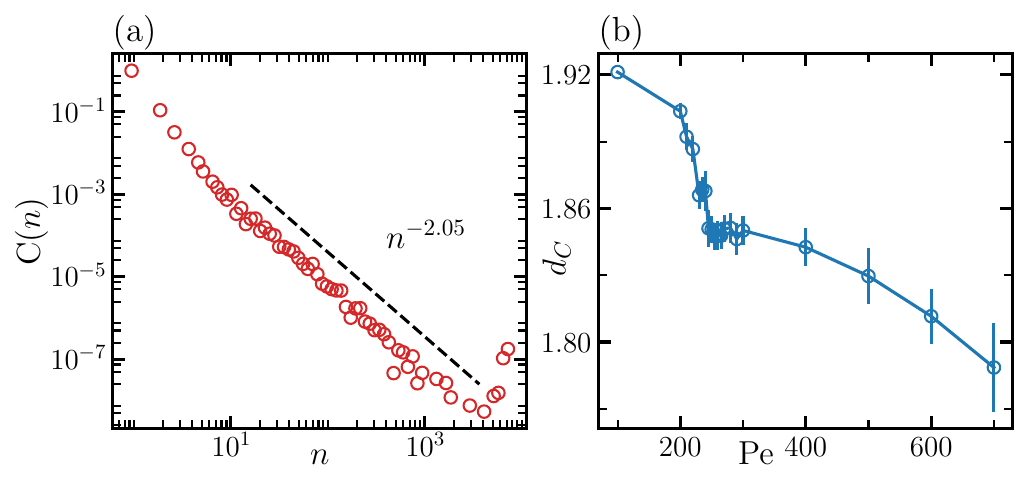}
\caption{(a) Cluster size distribution $C(n)$ at $\mbox{Pe} = 550$ showing a power-law decay with exponent $\tau \approx -2.05$. (b) Correlation dimension ($d_C$), measured inside the largest cluster for systems with $f_L > 0.4$, as a function of Pe. }
\label{fig3}
\end{figure}

Percolation transitions are also associated with a scale-free size distribution of the clusters.  We therefore calculate the cluster-size distribution $C(n)$ near the transition point in region III. As shown in Fig.\ref{fig3}(a), we see a clear power-law behaviour, extending over several decades of length-scale. We find that $C(n)\sim n^{-\tau}$, where $\tau \approx 2.05$. This value is also consistent with the predictions for percolation transitions in 2D \cite{stauffer2018introduction,gawlinski1981continuum}. {Note that the peak seen at large $n$ is due to finite system size which scales as $N$ (see Fig.11 in Appendix-D).} 

We also quantify the porosity of the clusters at high Pe regions (Pe $\geq 100$) by calculating the correlation dimension ($d_C$) of the largest cluster~\cite{GrassbergerPRL1983,LAHMIRI2021,strogatz2018nonlinear}. The correlation dimension ($d_C$), being closely related to the fractal dimension \cite{strogatz2018nonlinear}, provides the information of the change in porosity of the largest cluster with change in particle motility.  At Pe$~\approx 100$, the system has already formed a dense cluster due to MIPS where we obtain $d_C \approx 1.95$. Upon increasing the Pe, the MIPS cluster in the MIPS state becomes porous near its surface, leading to a sharp decrease in $d_C$, as shown in Fig.\ref{fig3}(b). With further increase in Pe its internal structure becomes porous, which leads to a monotonic decrease in $d_C$.

\begin{figure}
\centering
\includegraphics[width=0.5\textwidth]{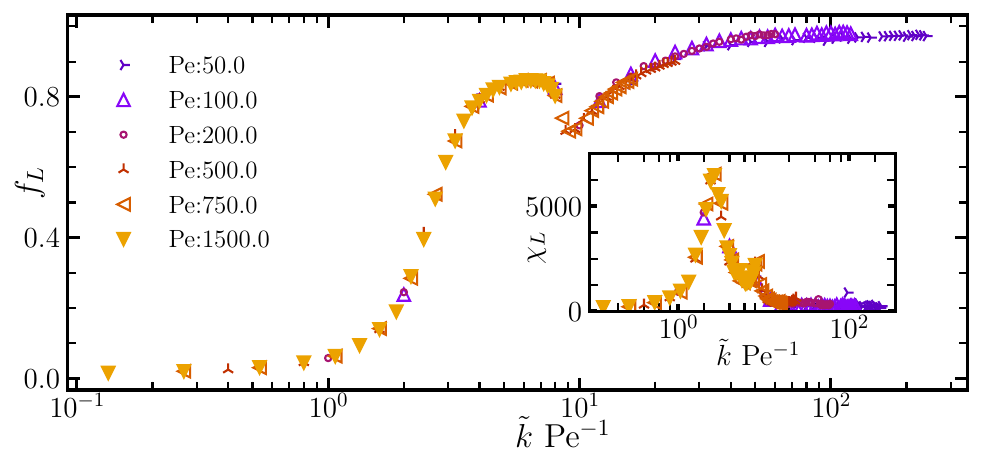}
\caption{ The largest cluster fraction $f_L$ as a function of stiffness parameter $\tilde{k}$ for different Pe values. Scaling the $\tilde{k}$ axis with $\text{Pe}^{-1}$ causes the data to collapse into a single curve. The corresponding scaling for susceptibility ($\chi_L$) is shown in the inset.}
\label{fig4}
\end{figure}
\subsection{Effect of interaction softness}
Finally, we study the effect of interaction softness on the phase properties by varying the stiffness constant $\tilde{k}$ keeping Pe constant. In Fig.\ref{fig4} we plot $f_L$ and $\chi_L$ (inset) as a function of $\tilde{k}$ for different values of Pe. Interestingly, the transitions observed earlier in region-II and III, with variation of Pe, are seen here as well but in the reverse order, when $\tilde{k}$ is increased. For each Pe we also see the formation of MIPS when $\tilde{k}$
is sufficiently large.  Interestingly, when we scale the $\tilde{k}$ axis with Pe$^{-1}$, the $f_L$ plots for different Pe collapse to a single curve (see Fig.~\ref{fig4}). The same scaling applies for $\chi_L$ as shown in the inset of Fig.~\ref{fig4}. A similar scaling also applies for the $f_L$ vs Pe curve for different $\tilde{k}$ values, as expected (see Fig.12 Appendix-E). Thus it is evident that an increase in Pe effectively makes the inter-particle interactions softer and an increase in motility is equivalent to decreasing the interaction stiffness. This scaling behavior can be further understood in terms of the effective distance $r_0$ between the particles \cite{sanoriaPRE2021}. 
When $\tilde{k}$ is decreased, the softening of interaction will lead to a linear decrease in $r_0$. We numerically estimate $r_0$ for different Pe from the distribution of interparticle distances $r_{ij}$ for pairs of interacting particles. The computed values of $r_0$ shows that in the limit of both high and low Pe, an increase in Pe will also cause a similar linear decrease in $r_0$ (see Fig.~13(c) in Appendix-F). Within the MIPS phase ($\mbox{Pe}\tilde{k}^{-1} \ll 1$), an approximate estimate shows that this linear dependence is of the form $r_0 \approx [1-{(C\tilde{k})}^{-1}~\mbox{Pe}]$, ($C$ is a constant, see Appendix-F) \cite{sanoriaPRE2021}. 
We also calculate the cutoff distance $r_c$ below which $p(r_{ij})$ is negligible (see Appendix-F). We find that $r_c$ also follows a similar linear decrease with Pe, similar to $r_0$, in both high and low Pe limits (Fig.~13(c) in Appendix-F). We also note that even for $\text{Pe} \approx 2\text{Pe}_c$ there is a significant non-zero excluded distance for a pair of interacting particles as verified directly from the simulations, indicating that the particles do not pass through each other in region III. 

\section{Summary and outlook}
In summary, we show that the softness plays a crucial role in determining the phase behaviour of active particles, which appears to be much richer than what was conceived so far.  It was believed that the ordered state of active Brownian particles with short-ranged repulsive interactions is relatively simple, where an increase in particle motility leads to a density ordering in the system and the destruction of this ordered state was attributed to particle crossing~\cite{fily2014freezing}. However, other active matter systems with orientational ordering have shown more complex density effects, such as coarsening \cite{DasDey2012} and percolation transitions \cite{kyriakopoulos2019clustering}. Here we have shown that complex structural effects can also be observed in systems without orientational ordering, purely driven by high particle motility and softness.
We have characterized the transition at high motility, by combining multiple quantities, namely the mean largest cluster fraction $f_L$, the linear extension of the largest cluster $d_L$, the susceptibility $\chi_L$, spanning probabilities $S_{x/y}$. We performed finite-size scaling of these quantities to confirm that the transition is a standard percolation transition.  
We have shown that during this particular transition, motility acts as an effective softness parameter and the structural properties can be scaled with $\mbox{Pe}^{-1}$. Therefore, the structural transitions that we observe is a softness induced one. When Pe is small, there exists another porous state, which disappears in the presence of translational noise. This porous network, unlike the percolated structure at large Pe, has small domains of high local hexatic order.


We note that the specific scaling form, $\text{Pe}^{-1}\tilde{k}$, is a consequence of the harmonic interaction potential, causing a linear decrease in minimum interparticle distance within the clusters, with increase in motility. Deviations from this scaling form are expected if a different form of interaction potential is used, for e.g., in many earlier studies, the potential diverges as $r \rightarrow 0$ \cite{Redner2013,digregorio2018full,sanoriaPRE2021}. Although we use a non-diverging potential, we observe that finite excluded region for interparticle distances vanishes only at very high motility  ($\text{Pe} > \text{Pe}_c$). Thus, we believe that the percolation transition observed at high motility is possibly a generic feature of active particles with soft interaction. However, these results are relevant for systems with rare particle crossings such as cells in tissues. Thus, we show that when we go beyond the idealized hard-particle interactions, new collective properties can arise in active Brownian systems. These findings are particularly important, since most of the natural systems does not consist of hard particles. Our study is also relevant in the context of recent interests in self assembly of soft colloidal particles~\cite{menath2021defined}.
 
\emph{Acknowledgement:}~The authors thank Dibyendu Das for insightful discussions. MS thank CSIR, India, for financial support.  We thank IIT Bombay HPC facility (Spacetime2).
\appendix
\section{Comparison of different structure}
\label{Appendix A}
\begin{figure*}
\centering
\includegraphics[width=0.8\textwidth]{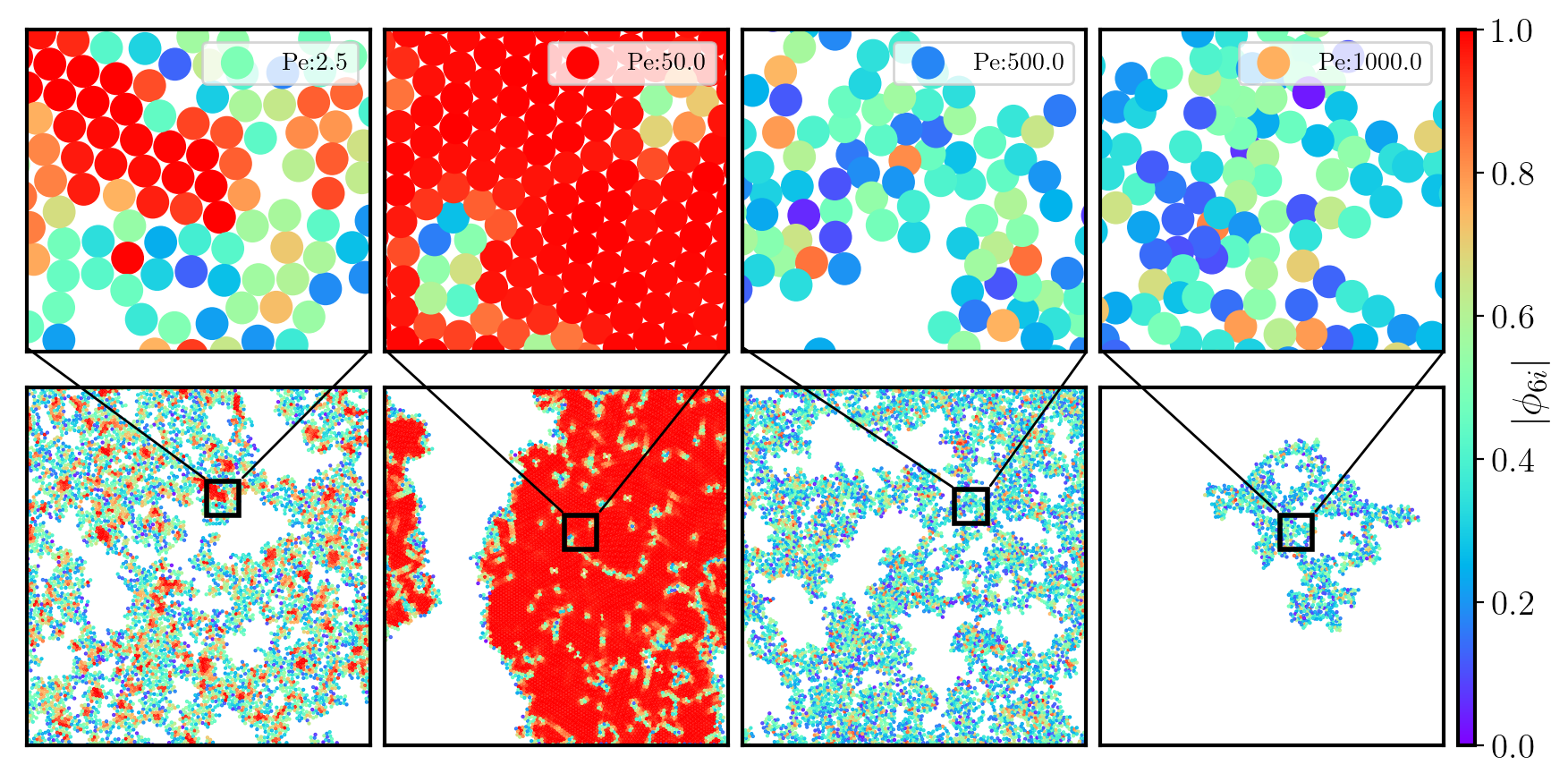}
\caption{Plots of steady state configurations for the entire system at different Pe values of 2.5, 50.0, 500.0, and 1000.0, as well as plots zoomed in on small sections; the colour of each particle shows the value of the local hexatic order $\left| \phi_{6i} \right|$}
\label{figSI9}
\end{figure*}

In Fig.~1 in the manuscript, we have compared the different phase space behavior, particularly the differences of the porous networks at high and low motility, as their difference with MIPS. Here we zoom into small scales to compare the structural differences. This is shown in Fig.~\ref{figSI9}.

\section{Detailed analysis of transitions}
\begin{figure}[h!]
\centering
\includegraphics[width=0.48\textwidth]{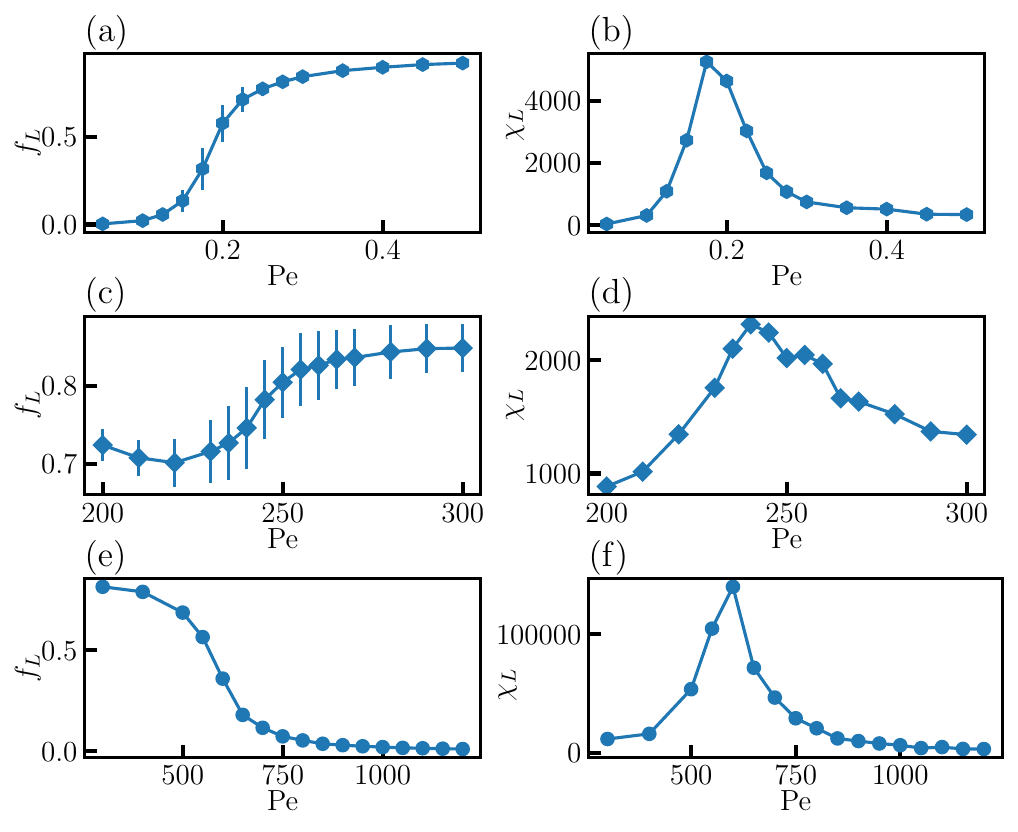}
\caption{Three transitions captured by the order parameter cluster fraction$f_L$ at fixed $\phi = 0.7$. Here, we show three regions in Pe, I(0.05-0.5), II(200-300), and III(550-1200) as shown in Fig.2(a) in main text separately. (a) The $f_L$ of the region I show the transition from a homogeneous region with small clusters to the largest connected cluster that spans the whole system as shown in its second cumulant plot (b).(c-d) shows the transition from MIPS to a percolated cluster of region II for box size $L = 210$. Similarly, (f-g) region III shows the percolation transition for box size $L = 1024$.}
\label{figSI2}
\end{figure}

\begin{figure}[h!]
\centering
\includegraphics[width=0.45\textwidth]{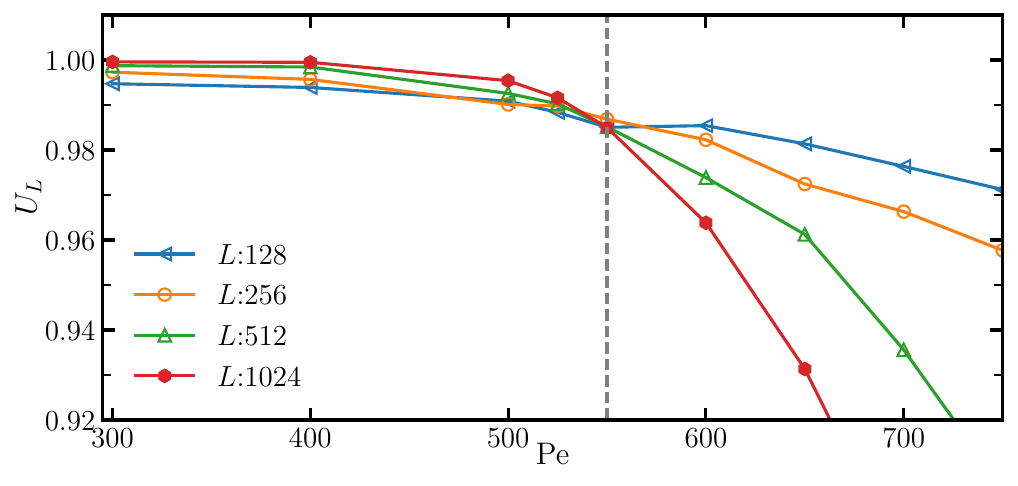}
\caption{The fourth order Binder cumulant $U_L$ is plotted as a function of Pe for four different systems sizes. The curves crosses at Pe $\simeq 550$.
}

\label{figSI3}
\end{figure}

We analyze the phase behaviour of active Brownian particles by calculating various  order parameters, averaged at the steady state. The fraction of largest cluster $f_L$ is one of the order parameters that can be used to detect the transitions. The corresponding susceptibility, which is defined for a system with size $L$, as $\chi_L = L^2 \sigma_L$, where $\sigma_L$ is the second cumulant of $f_L$ diverges at the transition (see inset of Fig.~2(a) in main text). Different transitional  regions are plotted separately in Fig.\ref{figSI2}. A reliable quantity that can be used to estimate transitions is the fourth order Binder cumulant for different system sizes $L$. The fourth order Binder cumulant {$U_L=\frac{1}{2}~(3 - \frac{\langle d_L^4 \rangle}{\langle d_L^2 \rangle^2})$} is calculated for the order parameter $d_L$ which is the normalized maximum linear extension of the largest cluster. The crossing point of $U_L$ for different $L$ provides $\mbox{Pe}_c\approx550$ (Fig.\ref{figSI3}).

\begin{figure}[h!]
\centering
\includegraphics[width=0.48\textwidth]{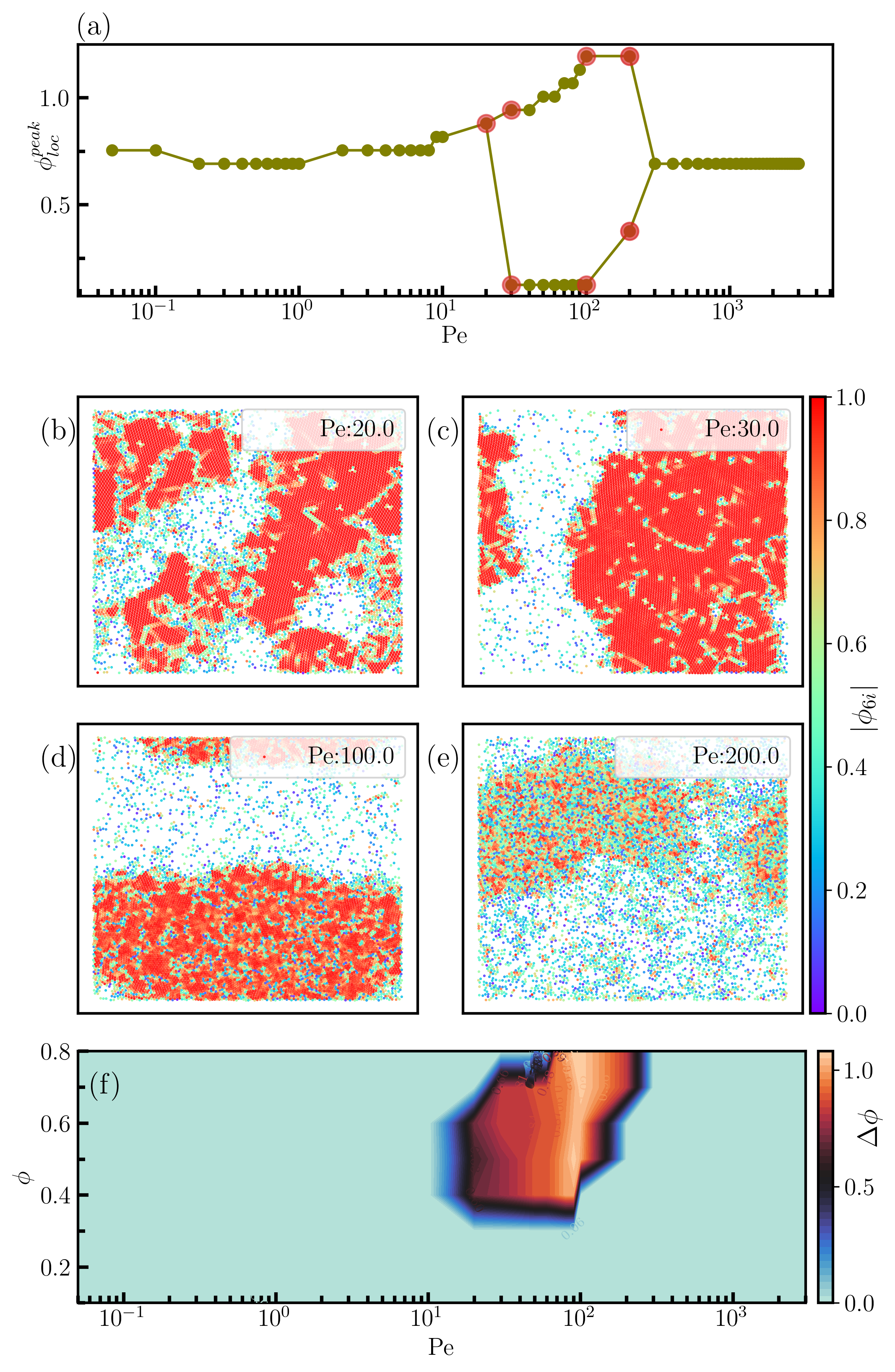}
\caption{ MIPS : (a)  The density corresponding to peak value of the local density distribution is plotted as a function of Pe at fixed $\phi = 0.7$. (b-e) are system configuration plots for the four distinct points $Pe = 20, 50, 100, 200$ here colour shows the value of local hexatic order $\left| \phi_{6i} \right|$  superimposed on particle configurations. (f) Phase diagram of the density difference $\Delta \phi = \phi_{loc}^{peak}(max) - \phi_{loc}^{peak}(min) $ between the dense and dilute regions (as shown in (a)) in the Pe-$\phi$ plane. A large $\Delta \phi$ indicates MIPS}
\label{figSI4}
\end{figure}

However, these order parameters are not helpful in identifying the phase separated region, which has been visually identified for $20 \gtrsim \text{Pe} \gtrsim 200$. To identify this state more quantitatively, we measure the local density $\phi_{loc}$ which shows a bimodal distribution when the system phase separates. In Fig.\ref{figSI4} we plot the location of peak values $\phi_{loc}^{max}$ of the distribution. This quantity is multi-valued in the phase separated region, hence an indicator for MIPS.


\section{The effect of translational noise}
\begin{figure}[h!]
\centering
\includegraphics[width=0.48\textwidth]{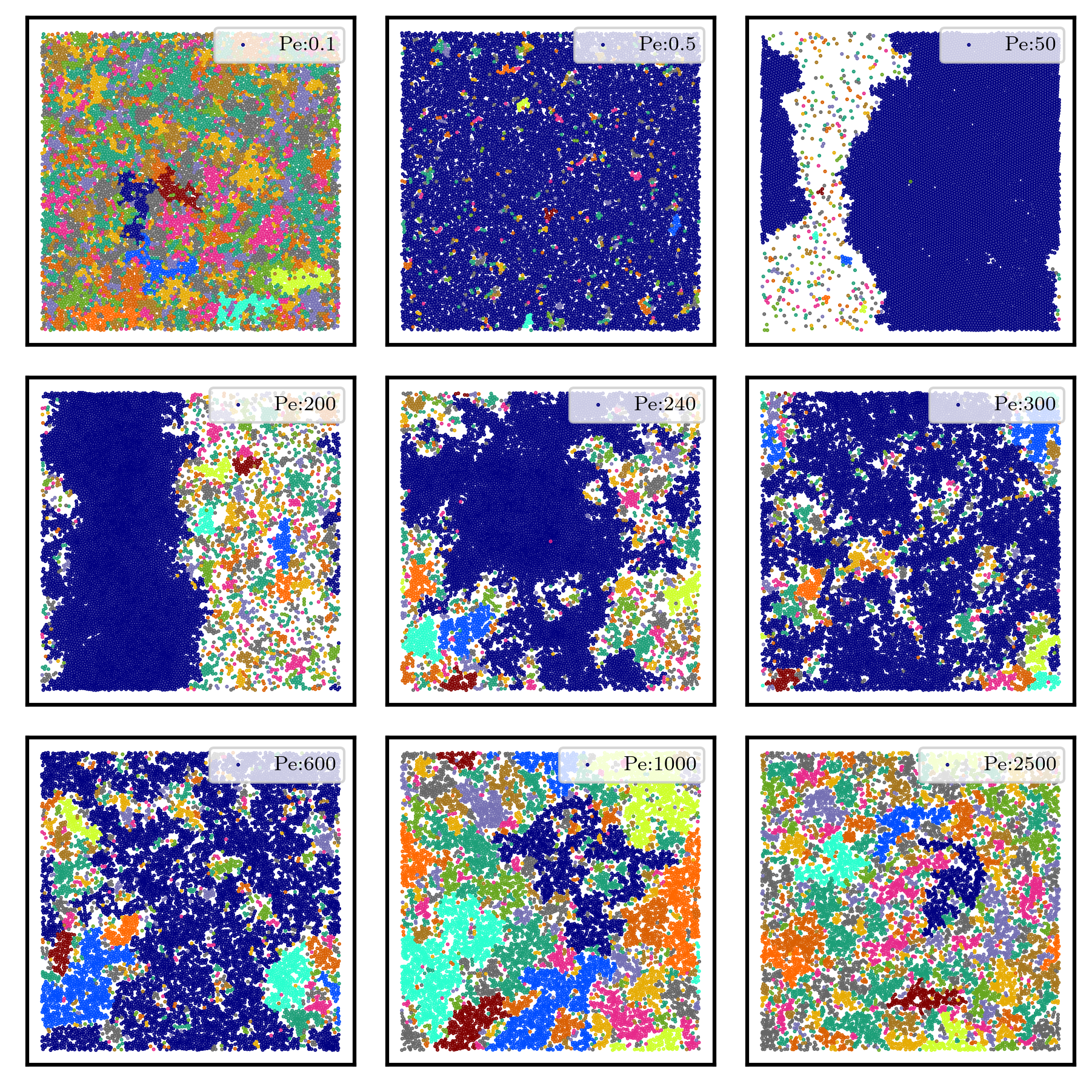}
\caption{Cluster distribution for Pe = 0.1, 0.5, 50, 200, 240, 300, 600, 1000, 2500.  Colour code blue indicates the largest cluster, and other colours represent the rest of the connected clusters. Due to a large number of various clusters, each colour is used for several separate clusters.}
\label{figSI5}
\end{figure}
\begin{figure}[h!]
\centering
\includegraphics[width=0.5\textwidth]{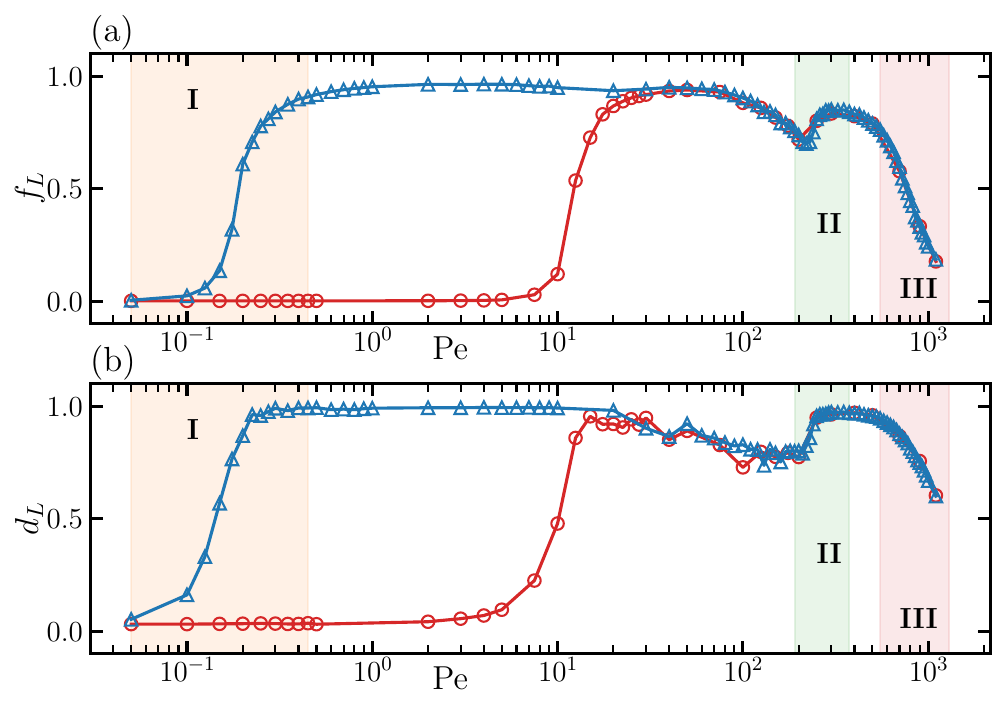}
\caption{Comparing system with and without translational diffusion: (a) mean largest cluster fraction $f_L$ as a function of the Pe at a constant $\phi = 0.7$ (b) normalized average linear extension $d_L$ as a function of Pe. In these plots blue curve represents the system without translational noise $D_t$ and red curve for system with $D_t$ and these are divided into three separate regions in Pe, I(0.05-0.5), II(200-300), and III(550-1100) corresponding to different transition regions as shown in Fig.2a of main text and Fig.\ref{figSI2} of SI.}
\label{figSI6}
\end{figure}
In the original dynamical equation for the particle positions (Eq.1 in the main text) the translational noise is not included. Although the effect of noise is expected to be negligible in the limit of large motility, it might influence the overall structural properties at low Pe. To study this effect, we run the simulations with an additional noise term to Eq(1),
\beqr
\label{eq1}
{\dot{\bf r}_i} &=& \mu~\sum_j{\bf F}({\bf r}_{ij})+ v_p~\hat{\mathbf{n}}_i+{\xi}_i,\\
\label{eq2}
\eqnr
where $\xi_i$ is the noise term which follows the relation $\left<\xi_i(t)~\xi_j(t')\right> = 2~D \delta_{ij}~\delta(t-t')$ and $\left< \xi_i \right>=0$. Here we choose $D=\frac{\sigma^2 D_r}{3}$. The structural properties of this system is analyzed by calculating $f_L$ and $d_L$ as a function of Pe, for $\phi = 0.7$. In Fig\ref{figSI6}, we compare these quantities with and without translational noise. Both $f_L$ and $d_L$ are almost same in both cases at large Pe (Pe $\gtrsim$ 20) beyond the MIPS region. However, at low Pe both $f_L$ and $d_L$ are qualitatively different compared to the athermal case. In the presence of noise, we no longer observe the formation of porous clusters at low Pe.

\section{Finite-sized effect in the Cluster size distribution}
\begin{figure}[h!]
\centering
\includegraphics[width=0.5\textwidth]{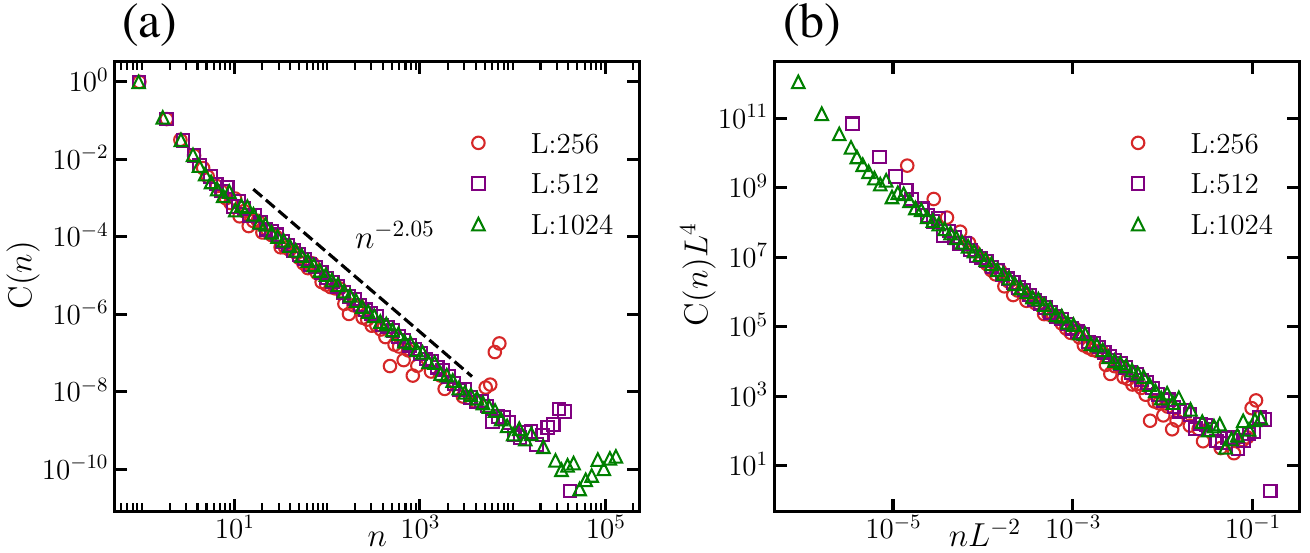}
\caption{(a) Cluster size distribution plot for our study shown for three different system sizes, namely $L=256, 512$, and 1024. (b) $x-$axis scaled by $L^{-2}$ to show the system size dependence.}
\label{fig5}
\end{figure}

In Fig.3(a) of the maintext, we see a peak at large $n$. This peak is a manifestation of finite system size.
To verify this, we have now plotted together the cluster size distribution for three different system sizes, namely $L=256,~512,$ and 1024 (see Fig.~\ref{fig5}(a)). As can be seen clearly, the peak gets shifted to larger $n$ values as $L$ is increased, while the power-law exponent remains unchanged. We also show in Fig.~\ref{fig5}(b) that if the $x$-axis is scaled by $L^2 \propto N$, the large cluster-size peaks collapses.

\begin{figure}[h!]
\centering
\includegraphics[width=0.48\textwidth]{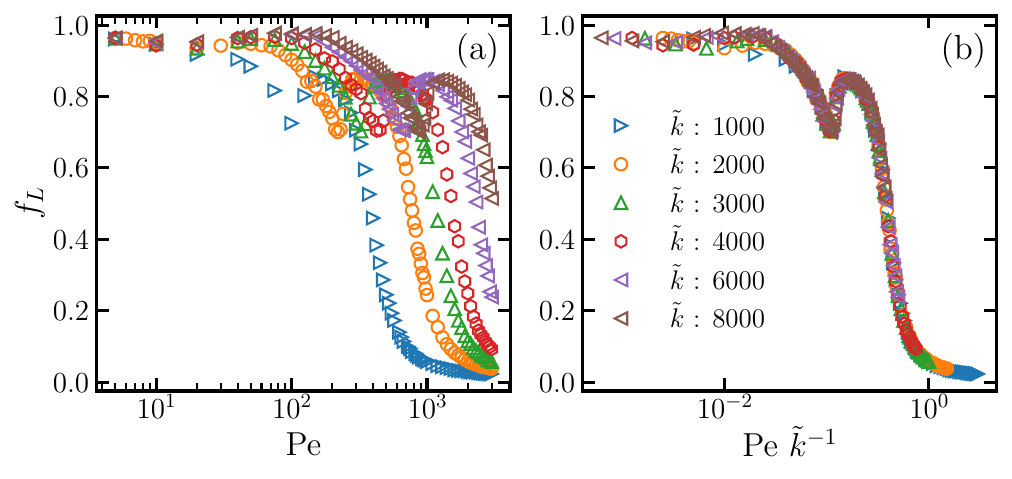}
\caption{(a) $f_L$ plotted as a function of Pe for different values of $\tilde{k}$. For $\mbox{Pe}<20$ the $f_L$ values are overlapping for all $\tilde{k}$. However,  there is a shift towards higher Pe with increase in $\tilde{k}$, for $\mbox{Pe}>20$. (b) the scaling of Pe with $\tilde{k}^{-1}$ collapses $f_L$.}
\label{figSI7}
\end{figure}
\section{Effect of interaction softness}
The results in Fig.2 in the main text shows the system behaviour for a fixed value of the stiffness parameter, $\tilde{k}=2000$. Here, we systematically study the effect of  $\tilde{k}$ in $f_{L}$ as a function of Pe (Fig.~\ref{figSI7}(a)). It is clear that at low Pe (Pe $\lesssim$ 20), the change in $\tilde{k}$ does not have a significant influence in $f_L$. However, where $Pe \gtrsim 20$, the interparticle interaction becomes crucial and we observe a systematic shift in the values of $f_L$ to higher Pe, when $\tilde{k}$ is increased. As expected, we these data points collapse by scaling the $x$ axis with  $\tilde{k}^{-1}$ (Fig.\ref{figSI7}(b)).
\begin{figure}[h!]
\centering
\includegraphics[width=0.48\textwidth]{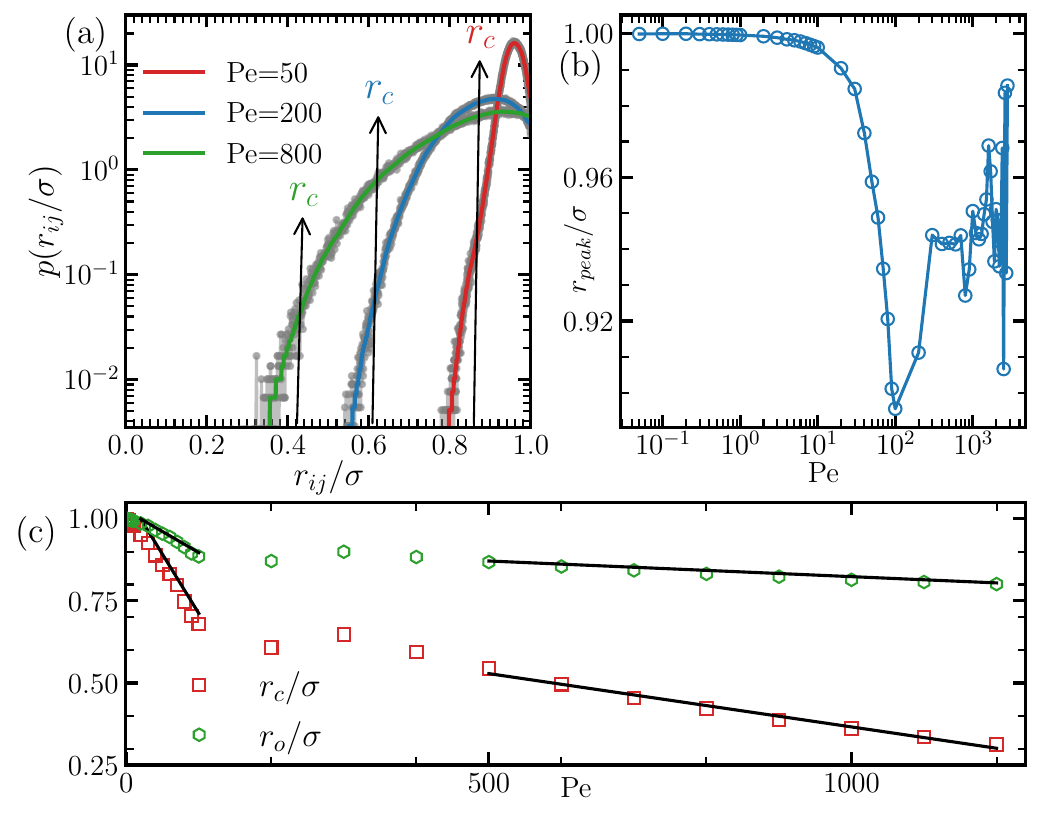}
\caption{(a) Interparticle distance distribution for Pe = 50,200, and 800, where $r_c/\sigma$ is cut off distance for which $p(r_{ij}/\sigma)$ is $1 \%$ of the peak value. (b)Semi-log plot of $r_{peak}/\sigma$ as a function of Pe. (c)  $r_c/\sigma$ and $r_0/\sigma$ plotted as a function of Pe. Black lines shows the corresponding linear fits.}
\label{figSI8}
\end{figure}

\section{Calculation of peak and cutoff interparticle distances}

An increase in interaction softness is manifested as a decrease in interparticle distance between a pair of interacting particles. To quantify this, we calculate the distribution of interparticle distances $p(r_{ij})$ between a pair of particles within the largest cluster. As shown in Fig.\ref{figSI8}(a), the distribution has {a maximum  at $r_{ij} \lesssim \sigma$}. For a given stiffness ($\tilde{k} = 2000$), we find that the width of the distribution increases while the peak value ($r_{peak}$) shifts only marginally (Fig.\ref{figSI8}(a)-(b)), especially for $Pe > 300$ where the MIPS state is destroyed. This increase in width indicates a deviation from the hexatic crystalline order and a higher degree of disorder within the cluster. 
We define the cutoff distance $r_{c}$ as the separation at which $p(r_{ij})$ decays into one percent of its maximum value at $r_{ij} = r_{peak}$. 
We plot $r_c$ as a function of Pe (see Fig.\ref{figSI8}(c) and observe a linear decrease with Pe in the limit of both high and low values of Pe. The cutoff $r_c$ has a significant non-zero value even for very high motility as evident in Fig~\ref{figSI8}(a) and (c). Further, we did not observe a single event with $r_{ij} \leq 0.2$ for $\tilde{k}=2000$ even for $\text{Pe} = 2\text{Pe}_c$. These observations verify that despite the interaction softness, the particles do not pass through each other in the high motility percolated state and there exists a lower cutoff for $r_{ij}$ below which the particles do not approach (Fig~\ref{figSI9}).  

We also plot the average interparticle distance $r_0$, calculated from the numerical data. Similar to $r_c$, $r_0$ also shows a linear decrease in both low and high Pe limits (Fig~\ref{figSI8}(c)). An approximate theoretical estimation of the average interparticle distance $r_0$ can be made within the MIPS state~\cite{sanoriaPRE2021}, which is given as
\[
 r_0 \approx \frac{1}{2}  \left[ 1+ \sqrt{1- {4\over C}\left({Pe \over \tilde{k}}\right)} \right]
\]
We note that within the MIPS region, since $(Pe /\tilde{k}) \ll 1$, $r_0$ also follows a linear relation, $r_0 \approx 1- {1\over {C}}({Pe \over \tilde{k}})$. Taking $C\approx0.39$, our estimate for $r_0$ within the MIPS state is in qualitative agreement with the data (see Fig.\ref{figSI8}(c)). 

Interestingly, after the percolation transition, both $r_c$ and $r_0$ within the largest cluster still scales linearly ($r_c\sim -0.65\tilde{k}^{-1}~\mbox{Pe}$ and $r_0\sim -0.19\tilde{k}^{-1}~\mbox{Pe}$, see Fig.\ref{figSI8}(c)). The behaviors of both $r_c$ and $r_0$ are consistent with the scaling used to collapse the data in Fig.~4 in the maintext. 
This behaviour explains the overall linear scaling of $f_L$ and $\chi_L$ with {$\tilde{k} \mbox{Pe}^{-1}$ in Fig.\ref{figSI7} and Fig.4 in the main text}.
We note that this linear decrease in $r_c$ is likely to be consequence of the particular form of the interaction. Many of the previous studies on ABPs have used interaction potentials which diverge as ${1\over r_{ij}^\alpha}$, where $\alpha >1$. In such cases the scaling of $r_c$ with Pe is expected to be different. However, since the overall physical properties are determined by the overlap distance, we believe that the percolation transition at high Pe (or high softness) will be observed for sufficiently large motilities, irrespective of the form of interparticle interactions.
\bibliography{bibg}
 
\end{document}